\documentstyle{article}
\begin{document}
\begin{center}
{\large Hamiltonian structure of real Monge-Amp\`{e}re equations}\\[20mm]
{\large Y. Nutku}\\[2mm]
 T\"{U}B\.{I}TAK - Marmara Research Center \\
 Research Institute for Basic Sciences \\
 Department of Physics \\
 41470 Gebze, Turkey \\[5mm]
\end{center}
\vspace{3cm}
\def\yy{\rule[-3.5mm]{0.5pt}{30pt}}

   The variational principle for the real homogeneous Monge-Amp\`ere
equation in two dimensions is shown to contain three arbitrary functions
of four variables. There exist two different specializations of this
variational principle where the Lagrangian is degenerate and furthermore
contains an arbitrary function of two variables. The Hamiltonian formulation
of these degenerate Lagrangian systems requires the use of Dirac's theory of
constraints. As in the case of most completely integrable systems the
constraints are second class and Dirac brackets directly yield the
Hamiltonian operators. Thus the real homogeneous Monge-Amp\`{e}re equation
in two dimensions admits two classes of infinitely many Hamiltonian
operators, namely a family of local, as well as another family nonlocal
Hamiltonian operators and symplectic 2-forms which depend on arbitrary
functions of two variables. The simplest nonlocal Hamiltonian operator
corresponds to the Kac-Moody algebra of vector fields and functions
on the unit circle. Hamiltonian operators that belong to either class
are compatible with each other but between classes there is only one
compatible pair. In the case of real Monge-Amp\`{e}re equations with
constant right hand side this compatible pair is the only pair of Hamiltonian
operators that survives. Then the complete integrability of all these real
Monge-Amp\`{e}re equations follows by Magri's theorem. Some of the remarkable
properties we have obtained for the Hamiltonian structure of the real
homogeneous Monge-Amp\`{e}re equation in two dimensions turn out to be
generic to the real homogeneous Monge-Amp\`{e}re equation and the geodesic
flow for the complex homogeneous Monge-Amp\`{e}re equation in arbitrary
number of dimensions. Hence among all integrable nonlinear evolution
equations in one space and one time dimensions, the real homogeneous
Monge-Amp\`{e}re equation is distinguished as one that retains its character
as an integrable system in multi-dimensions.

\vspace{2cm}

\section{Introduction}

   Recently we had found \cite{ns} that the real homogeneous
Monge-Amp\`{e}re equation (RHMA) in two dimensions, hereafter to
be referred to as RHMA$_2$, admits generalized Hamiltonian structure.
In field theory this type of structure arises for degenerate Lagrangian
systems where we must use Dirac's theory of constraints \cite{dirac} to
cast the problem into Hamiltonian form and finds its zenith in the theorem
of Magri \cite{magri} for completely integrable bi-Hamiltonian systems.
We refer to refs. \cite{mmt} for an exposition of this subject.
In this paper we shall show that RHMA and the geodesic flow for the 
complex homogeneous Monge-Amp\`{e}re equation (CHMA) admit infinitely many
Hamiltonian structures in arbitrary dimension. By Magri's theorem this result
shows the complete integrability of RHMA and, in turn, RHMA provides the
richest illustration of Magri's theorem.

    It will be useful to first list the various homogeneous Monge-Amp\`{e}re
equations that we shall discuss in order to fix the notation.
On a real manifold $M$ of dimension $n$, RHMA$_n$ is given by
\begin{equation}
\;\;\;\;\;\;\;\;\; \det u_{\alpha \beta} = 0     \;\;\;\;\;\;\;\;\;\;  
           \alpha = 0,1,...,n-1
\label{ma}
\end{equation}
where
$$ u_{\alpha \beta} \equiv \frac{\partial^2 \; u \;}
           {\partial x^{\alpha} \partial x^{\beta} } $$
is the matrix of second derivatives. We shall take $x^0 = t$ and
write eq.(\ref{ma}) as a first order system of nonlinear evolution
equations. Without loss of generality we may assume
\begin{equation}
 \det u_{a b} \ne 0 \, ,     
\;\;\;\;\;\;\;\;\;\;  a = 1,...,n-1
\label{nondeg}
\end{equation} 
which is a statement of the nondegeracy of eq.(\ref{ma}).
In $ 1+1$ dimensions we have the simplest case of real Monge-Amp\`{e}re
equations
\begin{equation}  \begin{array}{rcl}
u_{tt} u_{xx} - u_{tx}^{\;\;\;2} & = & - K \, ,  \\
                               K & = & \pm 1, \; 0
\end{array}
\label{rhma2}
\end{equation}
which is hyperbolic, elliptic, or homogeneous (RHMA$_2$) respectively.
We shall show
that there exist two types of Hamiltonian operators for RHMA$_2$, a
family of local (\ref{jmu}), as well as nonlocal (\ref{jfk}) operators,
both of which contain an arbitrary function of two variables.

   We shall start with a discussion of the variational principles underlying
RHMA$_2$. The Lagrangian which results in the equations of motion
for RHMA$_2$ contains three arbitrary functions of four variables. 
We shall concentrate on two qualitatively different classes of Lagrangians 
which are specializations of the master Lagrangian (\ref{varpr})
for RHMA$_2$, {\it c.f.} eqs.(\ref{lmu})
and (\ref{lagpq}) below, both of which contain arbitrary functions and
which are furthermore degenerate. That is, the passage to a Hamiltonian 
formulation of these degenerate Lagrangian
systems gives rise to second class constraints 
as in the case of most completely integrable systems \cite{yn}.
The resulting two families of Hamiltonian operators,
{\it c.f.} eqs.(\ref{jmu}) and (\ref{jfk}) contain all the
information in the Dirac brackets of RHMA$_2$.
These Hamiltonian operators are infinite in number as they contain
arbitrary functions which make their first appearance in
the Lagrangian formulation of RHMA$_2$ which, in turn, is related to
its character as a universal field equation \cite{fgm}.
Any pair of local, or nonlocal Hamiltonian operators are compatible,
however a local Hamiltonian operator is not in general compatible with 
a nonlocal Hamiltonian operator with the exception of $J_0$ of
eq.(\ref{jmu01}) and $J_{1}$ given by eq.(\ref{j2pq}).
These two Hamiltonian operators determine part of the
multi-Hamiltonian structure of the real Monge-Amp\`{e}re
equation (\ref{rhma2}) when the right hand side is a non-zero constant.

    We shall show that the Kac-Moody algebra that corresponds to the
simplest nonlocal Hamiltonian operator (\ref{j2pq}) for RHMA$_2$
consists of the algebra of vector fields and functions on $S^1$.
The family of local Hamiltonian operators can be brought to the
standard form of a first order operator by a Miura transformation.
The recursion operator obtained by a composition of the simplest local
and nonlocal Hamiltonian operators is Sheftel'-type \cite{sh}, that is, it
can be written as the square
of a first order operator which is a good recursion operator itself.
Finally, we shall present the symplectic structure of RHMA$_2$
which is dual to these Hamiltonian structures. We shall also show
that the symplectic 2-forms obtained in this way are the time
components of the Witten-Zuckerman \cite{cwz} 2-form that is the main
geometrical object in the covariant formulation of symplectic structure.

     In the theory of developable surfaces there is an equation which
may be called Ur-RHMA$_2$
\begin{equation}
u_t \, u_x = k
\label{urma}
\end{equation}
where $k$ is constant,
as it is well-known \cite{ch} that solutions of eq.(\ref{urma}) must
satisfy RHMA$_2$. We shall discuss the Hamiltonian structure of Ur-RHMA$_2$
in section \ref{sec-urma2} and show that the Hamiltonian operators
appropriate to this equation consist of a scalar version of the
Hamiltonian operators for RHMA$_2$ which lends further credence to its name.

     We shall show that RHMA is an example of a nonlinear evolution
equation that is integrable in {\it arbitrary} number of dimensions, see
\cite{fgm} for other universal field equations with this property.
In section \ref{sec-rhman} we shall find that some of the remarkable
results we shall report on the Hamiltonian structure of RHMA$_2$
hold for RHMA$_n$ as well.
Hence RHMA$_2$ will serve as the prototype for the discussion of the 
symplectic structure of all real homogeneous Monge-Amp\'{e}re equations. 
The Hamiltonian structure of the complex Monge-Amp\`{e}re equation is
quite different from that of RHMA and will not be considered in this paper.
On the other hand the Hamiltonian structure of the geodesic flow
defined by CHMA \cite{semmes}, which will be discussed in section
\ref{sec-gchma}, is very similar to that of RHMA.

\section{First order evolutionary form of RMA$_2$}

   For the Hamiltonian treatment of real Monge-Amp\`{e}re equations
we need to rewrite them as a system of nonlinear evolution equations which
is first order in time. For RHMA$_2$ this is most conveniently
accomplished by introducing the definitions
\begin{equation}
 u_{x}  = p\, , \;\;\;\;\;\;\;  u_{t}  = q
\label{upq}
\end{equation}
and accordingly the equations of motion are either given by
\begin{equation}
u_{t}  = q \, , \;\;\;\;\;\;\;\;\;\;      q_{t}  =
\frac{\textstyle{1}}{\textstyle{u_{xx}}} \left( q_{x}^{\;\;2} - K \right)
  \; ,
\label{uq}
\end{equation}
or
\begin{equation}
p_{t}  =  q_{x} \, , \;\;\;\;\;\;\;\;\;\;
q_{t}  =  \frac{\textstyle{1}}{\textstyle{p_{x}}} \left( q_{x}^{\;\;2} -
                  K \right)
\label{deffpq}
\end{equation}
but we note that this split of eq.(\ref{rhma2}) into either one of these
pair of evolution equations is not unique. Thus we have necessarily
introduced a degree of freedom over and above that present in
eq.(\ref{rhma2}) itself.
We shall return to this point in section \ref{sec-miura} on the
Miura transformation.

    Henceforth we shall use the notation
$ \{ u^i ; \; i = 1,2 \} $ for the first order variables
where $ u^1 $ will stand for either $ u $, or $ p $ and $ u^2 \equiv q $.
Then the vector field defining the flow for eq.(\ref{rhma2}) is given by
\begin{equation}
{\bf X}_u = q \frac{\partial}{\partial u} +
      \frac{\textstyle{q_{x}^{\,2} - K}}
           {\textstyle{ u_{xx}}} \frac{\partial}{\partial q}
                            , \hspace{1cm}
{\bf X}_p = q_x  \frac{\partial}{\partial p} +
      \frac{\textstyle{q_{x}^{\,2} - K}}
           {\textstyle{ p_{x}}} \frac{\partial}{\partial q}
\label{pq}
\end{equation}
respectively. We shall cast the equations of motion following from
eqs.(\ref{pq}) into the form of Hamilton's equations
\begin{equation}
u^{i}_{\,t} = {\bf X} ( u^i )  = \{ u^i , H \} = J^{ik} \; \delta_{k} H
\label{hameqpq}
\end{equation}
where the Hamiltonian operator $J$ defining the Poisson bracket
is a skew-symmetric matrix of
differential operators satisfying the Jacobi identities
and $\delta_i$ denotes the variational derivative with respect
to $u^i$.

\section{Variational formulation of RHMA$_2$}

   We shall find that the Hamiltonian operators for RHMA$_2$ are doubly
infinite in number. The reason for the existence of such a multitude
of Hamiltonian structures
can be traced back to the fact that the variational principle (\ref{varpr})
for the real homogeneous Monge-Amp\`ere equation contains arbitrary
functions. Our approach to the construction of the Hamiltonian operators as
Dirac brackets appropriate to degenerate Lagrangian systems will make
it manifest that the existence of arbitrary functions in Lagrangians
for RHMA$_2$ is responsible for their appearance in the Hamiltonian
operators.

   It can be directly verified that the equations of motion following
from the variational principle
$$ \delta I = 0\, ,   \; \;\;\;\;\;\; I = \int {\cal L} \; dt \, dx $$
with Lagrangian density
\begin{equation}      \begin{array}{lll}
{\cal L} & = & l_1 \; u_{tt} + l_2 \; u_{tx} + l_3 \; u_{xx} \; ,  \\[3mm]
l_{\alpha} & = & l_{\alpha}   \left( u_t , u_x ,
    \frac{\textstyle{u_{tt}}}{\textstyle{u_{tx}}} ,
    \frac{\textstyle{u_{tx}}}{\textstyle{u_{xx}}}
    \right)     \;\;\;\;\; \alpha = 1, 2, 3
\end{array}
\label{varpr}
\end{equation}
yield RHMA$_2$ in the form of eq.(\ref{rhma2}).
There are three arbitrary functions of four variables in this Lagrangian.
We have not explicitly indicated the possible dependence of $l_{\alpha}$
on the third combination of the ratio of
second derivatives above as it is derivable from the others.
It appears that this is the richest example of a variational principle
for any nonlinear partial differential equation.
Fairlie, Govaerts and Morozov \cite{fgm} have considered the hierarchy of
field equations where the equations of motion at any level are proportional
to the Lagrangian at the next. They pointed out that this is a finite
hierarchy ending at ``universal field equations." RHMA$_2$ is
precisely such a universal field equation and this is
the reason why we should expect arbitrary functions in the
Lagrangian (\ref{varpr}).

     In the variational formulation
of eqs.(\ref{uq}) we shall consider various specializations of the
arbitrary functions $l_{\alpha}$ in eq.(\ref{varpr})
which will still contain arbitrary functions
of a more specialized variety. It is, however, important to remember
that the enormous number of possibilities offered by the master Lagrangian
(\ref{varpr}) is the source of all the results we shall present below.

\section{Local Dirac bracket for RHMA$_2$}
\label{sec-locdir}

    A useful specialization of the Lagrangian density (\ref{varpr}) that
results in the first order equations of motion (\ref{uq})
for RHMA$_2$ is given by
\begin{equation}
{\cal L}_{\lambda} = \lambda \, q_{t}
- \lambda_{u_{x}} \left( u_{t}  - q \right) q_{x}
\label{lmu}
\end{equation}
where
\begin{equation}
\lambda = \lambda ( u_x , q ) ,     \;\;\;\;\;\;\;\;
 \lambda_{u_{x} u_{x}} \ne 0
\label{lam2n0}
\end{equation}
is a twice differentiable arbitrary function of two variables.
The canonical momenta appropriate to this Lagrangian are given by
\begin{equation}  \begin{array}{llllll}
 \pi_1   \equiv &    \pi_{u} &
  = & \frac{\textstyle {\partial
 {\cal L}_{\lambda}  } }
 {\textstyle {\partial u_t} } &
 = & - \lambda_{u_{x}}  q_{x}  \, ,  \\[4mm]
 \pi_2   \equiv &    \pi_{q} &
 = & \frac{\textstyle {\partial
 {\cal L}_{\lambda}  } }
 {\textstyle {\partial q_t} }  &
 = &  \lambda   \end{array}
\label{momenta}
\end{equation}
subject to the canonical Poisson bracket relations
\begin{equation}
 \{ \pi_i (x) , u^k (y) \}  = \delta^{i}_{k} \, \delta ( x - y)
\label{canonical}
\end{equation}
with all others vanishing.
But the momenta (\ref{momenta}) cannot be inverted for the velocities, or
alternatively the Hessian vanishes
\begin{equation}
 det \;\; \yy \;
 \frac{\partial^2 {\cal L}_{\lambda}
 }{\partial u^{i}_{\,t} \; \partial u^{k}_{\,t}} \; \yy \; = 0
\label{hessianl}
\end{equation}
and we have a degenerate Lagrangian system. Thus the passage to
the Hamiltonian formulation of the Lagrangian (\ref{lmu}) requires
the use of Dirac's theory of constraints.

   Following Dirac we introduce the two primary constraints that result
from eqs.(\ref{momenta})
\begin{equation}  \begin{array}{ll}
\phi_1 = & \pi_{u} +  \lambda_{u_{x}}  q_{x} \,  , \\[2mm]
\phi_2 = &  \pi_{q} -  \lambda   \end{array}
\label{constraints}
\end{equation}
and calculate their Poisson brackets using the canonical Poisson
bracket relations (\ref{canonical}). The result
\begin{equation}  \begin{array}{lll}
 \{\phi_1(x), \phi_1(y)\}  & = & q_{y} \, \lambda_{u_{y} u_{y}} \,
        \delta_{y} (x-y)
      - q_{x} \, \lambda_{u_{x} u_{x}} \, \delta_{x} (y-x) \, ,  \\[2mm]
 \{\phi_1(x), \phi_2(y)\} & = &  \lambda_{u_{y}} \,
 \delta_{y} (x-y) - \lambda_{u_{x}} \, \delta_{x} (y-x)
  - q_{x} \, \lambda_{q u_{x}} \, \delta (y-x) \, ,  \\[2mm]
\{\phi_2(x), \phi_2(y)\} & = & 0
\end{array}
\label{pbconst}
\end{equation}
shows that the constraints (\ref{constraints}) are second class
as they do not vanish modulo the constraints. This is a
typical situation for integrable systems as in the example of
KdV, or shallow water equations \cite{yn}. I am grateful to
C. A. P. Galv\~{a}o \cite{capg} for pointing out to me that one
should not simplify the Poisson brackets of the constraints (\ref{pbconst})
using the rules for manipulating distributions as such
``simplifications" often lead to incorrect Dirac brackets.

     The total Hamiltonian of Dirac is given by
\begin{equation}
   H_{T} =  \displaystyle{\int}
\left( \pi_{i} \, u^{i}_{\;t} - \displaystyle{\cal L}
                   + c^{i} \phi_{i}    \right) dx
\label{ht}
\end{equation}
where $c^i$ are Lagrange multipliers and summation over $i=1,2$ is implied.
The condition that the constraints are maintained in time
\begin{equation}
  \{ \phi_i (x) , H_{T} \} = 0
\label{seco}
\end{equation}
gives rise to no further constraints which would have been secondary
constraints. Instead, using eqs.(\ref{pbconst}), we find that the
Lagrange multipliers are determined from eqs.(\ref{seco})
$$ c^1 = q \, , \hspace{2cm} c^2 = \frac{q_{x}^{\;\;2}}{u_{xx}} $$
and they do not depend on the choice of the arbitrary function $\lambda$.
This is expected since the constraints and therefore the total Hamiltonian
is linear in the momenta, the correct equations of
motion will result only if the Lagrange multipliers are simply the
components of the vector field (\ref{pq}) for the flow.
Finally from eq.(\ref{ht}) we find
\begin{equation}
 H_{\lambda \, T}
 = \int \left( q \, \pi_{u} + \frac{q_{x}^{\;\;2}}{u_{xx}} \, \pi_{q}
-  \frac{q_{x}^{\;\;2} \; \lambda }{u_{xx}} \right) d x
\end{equation}
for Dirac's total Hamiltonian. The check that with this total Hamiltonian all
the equations of motion are satisfied is straight-forward and we can
summarize all of them in Hamilton's equations
\begin{equation}
{\cal A}_{t} = \{ {\cal A} , H_{\lambda \, T} \}
\label{eqsofmo}
\end{equation}
where ${\cal A}$ is any functional of the canonical variables.
There is, however, one further simplification that we can carry out
because in Dirac's theory second class constraints hold as strong
equations. This fact gives us the choice of eliminating the momenta from
eqs.(\ref{momenta}). Then we can write
\begin{equation}
{\cal H}_{\lambda \, T} = q \, q_{x} \, \lambda_{u_{x}}
\end{equation}
for the total Hamiltonian density of Dirac.

   Given any two differentiable functionals of the canonical variables
${\cal A}$ and ${\cal B}$, the Dirac bracket is defined by
\begin{equation} \begin{array}{ll}
\{ {\cal A}(x), {\cal B}(y) \}_{D} = & \{ {\cal A}(x), {\cal B}(y) \} \\[4mm]
      & - \, \displaystyle{\int} \{ {\cal A}(x), \phi_i(z) \}
J^{ik}(z,w)  \{ \phi_k(w) , {\cal B}(y) \} \, dz \, dw
\end{array}
\label{defdirac}
\end{equation}
where $J^{ik}$ is the inverse of the matrix of Poisson brackets of the
constraints. The definition of the inverse is simply
\begin{equation}
\int \{\phi_{i}(x),\phi_{k}(z) \} J^{kj} (z,y) \; d z
                  = \delta_{i}^{j} \, \delta ( x - y )
\label{mik}
\end{equation}
which results in a set of differential equations for the
entries of $J^{ik}$.  Starting with the Poisson bracket relations
(\ref{pbconst}) we find that eqs.(\ref{mik}) can be solved readily to yield
\begin{equation}  \begin{array}{ll}
 J^{11}_{\lambda} (x, y) = & 0 \, , \\[3mm]
 J^{12}_{\lambda} (x, y) = & - \, \frac{\textstyle{1}}
 {\textstyle{ \lambda_{u_{x} u_{x}} \, u_{xx} }} \, \delta (x-y) \, , \\[5mm]
 J^{21}_{\lambda} (x, y) = &  \frac{\textstyle{1}}
 {\textstyle{ \lambda_{u_{x} u_{x}} \, u_{xx} }} \,\delta (x-y) \, , \\[5mm]
 J^{22}_{\lambda} (x, y) = & -  \frac{\textstyle{2 \, q_{x}}}
  {\textstyle{ \lambda_{u_{x} u_{x}} \, u_{xx}^{\;\;\;2} }} \,
                           \delta_{x} (x-y)
    - \left(  \frac{\textstyle{q_{x}}}
    {\textstyle{ \lambda_{u_{x} u_{x}} \, u_{xx}^{\;\;\;2} }}
    \right)_{\!x}   \, \delta (x-y)
\end{array}
\label{mm}
\end{equation}
for the inverse of (\ref{pbconst}).
It will be convenient to rename the arbitrary function $\lambda$ as
$\mu$ according to
\begin{equation}
\lambda_{u_{x} u_{x} } \equiv \frac{1}{q} \, \mu_{q}
\label{lamusi}
\end{equation}
for ease of calculations that will follow.

\section{Local Hamiltonian operators for RHMA$_2$}

   The transition from the Dirac bracket to the Hamiltonian operator
is given by
\begin{equation}
 \{ u^{i}(x) , u^{k}(y) \}_{D} = J^{ik}(x,y)  \equiv J^{ik}(x)  \delta(x-y)
\label{defhamop}
\end{equation}
and from eqs.(\ref{defhamop}) and (\ref{mm}) it follows that
the Hamiltonian operator corresponding to the
degenerate Lagrangian (\ref{lmu}) is simply
\begin{equation}
J_{\mu} =     \left(            \begin{array}{cc}
 0 & \frac{ {\textstyle  q} }{ {\textstyle
  \mu_{q} \,     u_{xx} } } \\[4mm]
- \frac{ {\textstyle q}}{  {\textstyle
   \mu_{q} \, u_{xx} } }
 &  \frac{ {\textstyle q \, q_x }}{{\textstyle
     \mu_{q} \, u_{xx}^{\;\;2} }} D_x
+ D_x  \frac{ {\textstyle q \, q_x }}{{\textstyle
    \mu_{q} \, u_{xx}^{\;\;2} }}
\end{array}  \right)               \label{jmuuq}
\end{equation}
which contains the arbitrary function $\mu$ of two variables.
Two particular choices of this arbitrary function, namely
$\mu = \frac{1}{2} q^{2}$ and $\mu = u_{x} q $ result in
the bi-Hamiltonian structure of eq.(\ref{rhma2}) reported in \cite{ns}.

   Under the change of variable $ p = u_x $ the Hamiltonian
operator (\ref{jmuuq}) is transformed
\begin{equation}
J_{\mu} =     \left(            \begin{array}{cc}
 0 &  D_{x}  \frac{ {\textstyle  q} }{ {\textstyle
  \mu_{q} \,     p_{x} } } \\[4mm]
 \frac{ {\textstyle q}}{  {\textstyle  \mu_{q} \, p_{x} } } D_{x}
 &  \frac{ {\textstyle q \, q_x }}{{\textstyle
     \mu_{q} \, u_{xx}^{\;\;2} }} D_x
+ D_x  \frac{ {\textstyle q \, q_x }}{{\textstyle
    \mu_{q} \, u_{xx}^{\;\;2} }}
\end{array}  \right)               \label{jmu}
\end{equation}
which is the form of the local Hamiltonian operator for RHMA$_2$
that we shall use henceforth. The most important element of the family
of Hamiltonian operators in eq.(\ref{jmu}) is given by
\begin{equation}
J_{0} =   \left(            \begin{array}{cc}
 0 & D_x \frac{\textstyle 1}{\textstyle{p_{x} } } \\[4mm]
 \frac{ {\textstyle 1}}{  \textstyle{p_{x}} } D_x
 &  \frac{\textstyle{ q_x }}{\textstyle{ p_x^{\;2} }} D_x
+ D_x  \frac{\textstyle{ q_x }}{\textstyle{ p_x^{\;2} }}
\end{array}  \right)               \label{jmu01}
\end{equation}
which corresponds to the choice  $ \mu =  \frac{1}{2} q^2 $.

   From the construction of the Dirac bracket in section \ref{sec-locdir} it
is clear that the reason for the presence of the arbitrary function
$\mu$ in the Hamiltonian operator (\ref{jmu}) can be traced back to the
degenerate Lagrangian (\ref{lmu}) where it makes its first appearance as
$\lambda$. In order to proceed we need to prove that (\ref{jmu}) is indeed
a Hamiltonian operator which requires a check of the Jacobi identities.
However, this also follows from the fact that (\ref{jmu}) is derived
from the Dirac bracket (\ref{mm}) for which we have a general proof of
the Jacobi identities \cite{pamd}.

\subsection{Jacobi identities}
\label{sec-jac23}

    The Hamiltonian operator is a bi-vector which defines the Poisson
bracket. The skew-symmetry of the operator  (\ref{jmu}) is manifest.
In order to verify that a given bi-vector is Hamiltonian
we must verify that it satisfies the tri-vector Jacobi identities.
Thus following Olver \cite{mmt} we introduce an arbitrary basis of tangent
vectors  $\Theta$ which are then conveniently manipulated according to the 
rules of exterior calculus.
The Jacobi identities are given by the compact expression
\begin{equation}
L \; \delta I = 0  \;\;\;\;\;\;\; ( mod. \; div. )
\label{jac}
\end{equation}
where
\begin{equation}
L = J \, \Theta, \;\;\;\;\;\;\;
 I = \frac{1}{2} \, \Theta^T \wedge J \Theta
 \label{ll}
\end{equation}
and $\delta$ denotes the variational derivative.
The vanishing of the tri-vector (\ref{jac}) modulo a
divergence is equivalent to the satisfaction of the Jacobi identities. 

    For the Hamiltonian operator (\ref{jmu}) we have a two component system
and introducing the basis
$$ \Theta = \left(  \begin{array}{c} \eta \\ \theta
\end{array}       \right) $$
we have from eqs.(\ref{ll})
\begin{equation}
L_{\mu}  = \left(  \begin{array}{c}
 \left( \frac{{\textstyle q}}{{\textstyle \mu_q p_x}} \right)_x   \theta
 +  \frac{{\textstyle q}}{{\textstyle \mu_q p_x}} \theta_x \\[4mm]
 \frac{{\textstyle q}}{{\textstyle \mu_q p_x}} \eta_x          +
2  \frac{{\textstyle q q_x}}{{\textstyle \mu_q p_x^{\;2} }}  \theta_x +
\left(  \frac{{\textstyle q q_x}}{{\textstyle \mu_q p_x^{\;2} }}
\right)_x \theta
\end{array}       \right) ,
\label{ljmu}
\end{equation}
$$ I_{\mu}  =  \frac{{\textstyle q }}{{\textstyle \mu_q p_x }}
\left( \theta \wedge \eta_x
 +  \frac{{\textstyle q_x }}{{\textstyle p_x }}
       \theta \wedge \theta_x \right) \, . $$
To form the required expression in eq.(\ref{jac}) we first calculate
the variational derivatives
\begin{equation}      \begin{array}{lll}
\frac{{\textstyle \delta I_{\mu} }}{{\textstyle \delta p}} & = & \left[
  \frac{{\textstyle q_x }}{{\textstyle \mu_q p_x^{\;2} }}
  -   \frac{{\textstyle q }}{{\textstyle \mu_q^{\;2} p_x^{\;2} }}
\left(  \mu_{qq} q_x + 2 \mu_{pq} p_x \right)
- 2  \frac{{\textstyle q p_{xx} }}{{\textstyle \mu_q p_x^{\;3} }}
\right] \theta \wedge \eta_x \\[5mm]
& & +   \left[  2
\frac{{\textstyle q_x^{\;2} + q q_{xx} }}{{\textstyle \mu_q p_x^{\;3} }}
- 2 \frac{{\textstyle \mu_{qq} q q_{x}^{\;2} }}{{\textstyle \mu_q^{\;2}
 p_x^{\;3} }}
- 3 \frac{{\textstyle \mu_{pq} q q_{x} }}{{\textstyle \mu_q^{\;2}
 p_x^{\;2} }}  - 6 \frac{{\textstyle q q_{x} p_{xx} }}{{\textstyle \mu_q
 p_x^{\;4} }} \right] \theta \wedge \theta_x  \\[5mm]
& & +
\frac{{\textstyle q }}{{\textstyle \mu_q p_x^{\;2} }}
\left( \theta_x \wedge \eta_x +  \theta \wedge \eta_{xx} \right)
+ 2 \frac{{\textstyle q q_x }}{{\textstyle \mu_q p_x^{\;3} }}
 \; \theta \wedge \theta_{xx} \, , \\[5mm]
\frac{{\textstyle \delta I_{\mu} }}{{\textstyle \delta q}} & = &  \left(
  \frac{{\textstyle 1 }}{{\textstyle \mu_q p_x }}     -
  \frac{{\textstyle q \mu_{qq} }}{{\textstyle \mu_q^{\;2} p_x }}
\right) \theta \wedge \eta_x     + \left(
  \frac{{\textstyle q \mu_{pq} }}{{\textstyle \mu_q^{\;2} p_x }} +
  2  \frac{{\textstyle q p_{xx} }}{{\textstyle \mu_q p_x^{\;3} }}
\right) \theta \wedge \theta_x   \\[5mm]
& & -  \frac{{\textstyle q }}{{\textstyle \mu_q p_x^{\;2} }}
 \; \theta \wedge \theta_{xx} 
\end{array}
\label{delijmu}
\end{equation}
and take their exterior product with $L_{\mu}$. The result
$$ L_{\mu}   \; \delta I_{\mu}  =
 -\left( \frac{{\textstyle q^2 }}{{\textstyle \mu_q^{\;2}
\;  p_x^{\;3} }} \; \theta \wedge \theta_x \wedge \eta_x \right)_x $$
is a total derivative so that the Jacobi identities are satisfied.

\subsection{Hamilton's equations}

   It can be directly verified that
\begin{equation}
{\cal H}_{\mu} = \mu \, p_{x}             \label{hmu}
\end{equation}
is conserved for the flow (\ref{pq}) and this is the Hamiltonian
density appropriate to the operator (\ref{jmu}). That is,
the equations of motion (\ref{pq}) are cast into the form of
Hamilton's equations (\ref{hameqpq}) with $J_{\mu}$ and $H_{\mu}$.
But since $\mu$ is an arbitrary function of $p$ and $q$,
given any other differentiable function $\nu = \nu(p,q)$ we have
\begin{equation}
 J_{\mu} \; \delta H_{\mu} =  J_{\nu} \; \delta H_{\nu}
\label{lm}
\end{equation}
which is a statement of the Lenard-Magri recursion relation.
Thus we have another expression of the fact that there exist
infinitely many Hamiltonian operators for RHMA$_2$.
The Casimir density for the Hamiltonian operator (\ref{jmu}) is given by
\begin{equation}
{\cal C}_{\mu}  = \sigma \, p_{x} \, ,
\label{casimir}
\end{equation}
where
\begin{equation}
\sigma_q = \frac{1}{q} \mu_q
\label{defsig}
\end{equation}
since it can be readily verified that $ J_{\mu} \delta C_{\mu} = 0$.
The proof of the Jacobi identities for (\ref{jmu}) in section \ref{sec-jac23}
is for arbitrary $\mu$, hence all local Hamiltonian operators $J_{\mu}$
are compatible with each other.

\subsection{Miura transformation}
\label{sec-miura}

 The remarkable feature of the Hamiltonian operator (\ref{jmu}) is the
existence of $\mu$, an arbitrary function of two variables.
We have already remarked that for the Hamiltonian formulation of
RHMA$_2$ we need to start with a pair of nonlinear evolution
equations and therefore had to use the equations of motion
defined by the vector field (\ref{pq}) rather than eq.(\ref{rhma2}) itself.
This introduces an extra degree of freedom as the definitions of
$p, q$ cannot be unique. It may appear that 
this extra degree of freedom could be responsible for the 
existence of the arbitrary function $\mu$ in the Hamiltonian operator 
(\ref{jmu}). This is not the case.
We have shown that the reason for the appearance of this arbitrary function
in the Hamiltonian operator can be traced back to the existence of
arbitrary functions in the master Lagrangian.
Nevertheless, there is certainly an extra degree of freedom
available in the definition of
the first order dependent variables for RHMA$_2$ and it is possible to
exploit it for various purposes. In fact by a redefinition of $q$ the
Hamiltonian operator (\ref{jmu}) can be brought to the standard form
of a first order operator with constant coefficients.

   In place of eq.(\ref{deffpq}) we may use the following
definitions for the auxiliary variables $p, q$
\begin{equation}
u_x = p , \;\;\;\;\; u_t = Q \left( p, \frac{q}{p_x} \right)
\label{deff2}
\end{equation}
where $Q$ is a differentiable function of its arguments.
The equations of motion are now given by
\begin{equation}
p_t = Q_x \; ,   \;\;\;\;\;\;\;\;
q_t = \left( \frac{q}{p_x} Q_x \right)_x
\end{equation}
which are already in standard Hamiltonian form. That is, we have
\begin{equation}   \left(
\begin{array}{l}  p_t \\ q_t \end{array} \right) =
 \left( \begin{array}{cc} 0 & D_x \\ D_x & 0 \end{array} \right)
 \left( \begin{array}{c}   \delta_p \\   \delta_q \end{array}
 \right)  H       \label{coef}
\end{equation}
with the familiar first order Hamiltonian operator and the Hamiltonian
density is given by
\begin{equation}
{\cal H} = h \left( p, \frac{q}{p_x} \right) p_x
\end{equation}
where $h$ is related to $Q$ through $ {\cal H}_q = Q $.
The transformation of variables obtained by a comparison of
eqs.(\ref{deff2}) and (\ref{upq}) is a Miura transformation
as it brings the Hamiltonian operator to the canonical form (\ref{coef})
of Gardner-Zakharov-Fadeev \cite{gzf} with constant coefficients.

\section{Nonlocal Dirac bracket for RHMA$_2$}

   There is another family of Hamiltonian operators for eqs.(\ref{uq})
which falls outside the class of Hamiltonian operators in eq.(\ref{jmu}).
Its origin can be traced back to the fact that
\begin{equation}
{\cal L}_{\kappa} =
      \kappa \,  p_x +  \kappa_{r} \left( p_{t} - q_{x} \right)
\label{lagpq}
\end{equation}
with $\kappa$ an arbitrary function of two variables
\begin{equation}
\kappa = \kappa \left( p, r \right) \, ,  \;\;\;\;\;\;
r \equiv \frac{q_x}{p_x} \, ,  \;\;\;\;\;\;  \kappa_{rr} \ne 0
\label{defkap}
\end{equation}
is a new Lagrangian for the system (\ref{uq}).
This is another specialization of the master Lagrangian
(\ref{varpr}) which is also degenerate and the passage to its
Hamiltonian formulation again requires the use of Dirac's theory.

   Starting with the Lagrangian (\ref{lagpq}) the momenta are given by
\begin{equation}  \begin{array}{llllll}
 \Pi_{1} \equiv & \Pi_{p}& = & \frac{
 \textstyle{\partial {\cal L}_{\kappa} } }
 {\textstyle{\partial p_t}}
& = &  \kappa_r \\[4mm]
 \Pi_{2} \equiv & \Pi_{q} & = & \frac{
 \textstyle{\partial {\cal L}_{\kappa} } }
                 {\textstyle{\partial q_t} }
& = &           0
\end{array}
\end{equation}
which cannot be inverted for the velocities. So we introduce the
primary constraints
\begin{equation}  \begin{array}{ll}
\Phi_1 = & \Pi_{p} - \kappa_r    \\[2mm]
\Phi_2 = &  \Pi_{q}
\end{array}
\label{constnl}
\end{equation}
and using the canonical Poisson bracket relations (\ref{canonical}) we find
\begin{equation}  \begin{array}{lll}
 \{\Phi_{1}(x), \Phi_{1}(y) \}
  & = & \kappa_{rr} (x) \,
  \frac{\textstyle{q_x}}{\textstyle{p_{x}^{\;2}}} \,
  \delta_{x} (x-y)  -
  \kappa_{rr} (y)     \,
  \frac{\textstyle{q_y}}{\textstyle{p_{y}^{\;2}}} \,
  \delta_{y} (y-x)   \\[5mm]
 \{\Phi_{1}(x), \Phi_{2}(y) \}
 & = &  \kappa_{rr}(x)                            \,
  \frac{\textstyle{1}}{\textstyle{p_{x}} }        \,
  \delta_{y} (x-y)    \\[4mm]
 \{\Phi_{2}(x), \Phi_{2}(y) \} & = &  0
\end{array}
\label{pbconstnl}
\end{equation}
which show that the constraints are once again second class.
The total Hamiltonian consists of a sum of the free Hamiltonian
with a linear combination of the constraints as in
eq.(\ref{ht}) and the requirement that the constraints
(\ref{constnl}) are maintained in time, {\it c.f.} eqs.(\ref{seco}),
does not lead to secondary constraints but instead
determines the Lagrange multipliers
$$ c^1 = q_{x}  \, , \hspace{2cm} c^2 = \frac{q_{x}^{\;\;2}}{p_{x}} $$
which also do not depend on the arbitrary function $\kappa$.
Using this information we find that
\begin{equation}
 H_{\kappa \, T} = \int \left(      q_x \, \Pi_{p}
 + \frac{q_{x}^{\;\;2}}{p_{x}} \, \Pi_{q}
- \kappa \, p_{x} \right) d x
\end{equation}
is the total Hamiltonian and the equations of motion are given
by eq.(\ref{eqsofmo}). Once again we can simplify the total
Hamiltonian using the fact that second-class constraints (\ref{constnl})
hold as strong equations and we find
\begin{equation}
{\cal H}_{\kappa \, T} = \kappa_{r} \, q_{x}  \,  -  \, \kappa \,  p_x
\label{hd}
\end{equation}
for the Hamiltonian density of Dirac.

   For the Dirac bracket we need the inverse of the Poisson brackets of
the constraints (\ref{pbconstnl}). The solution of eqs.(\ref{mik})
is given by
\begin{equation}  \begin{array}{ll}
 J^{11}_{\kappa} (x, y) = & 0 \, ,  \\[2mm]
 J^{12}_{\kappa} (x, y) = & - \frac{\textstyle{p_x}}
 {\textstyle{\kappa_{rr}(x) }} \theta (x-y) \, ,   \\[5mm]
 J^{21}_{\kappa} (x, y) = &  - \delta (x-y)  \displaystyle{\int}^x
 \frac{\textstyle{p_{\xi}}}{\textstyle{\kappa_{rr} (\xi) }} \,
                                                 d \xi \, , \\[6mm]
 J^{22}_{\kappa} (x, y) = &    -
 \frac{\textstyle{q_x}}{\textstyle{\kappa_{rr}(x) }} \theta (x-y)    -
    2 \, \delta (x-y)  \displaystyle{\int}^x
 \frac{\textstyle{q_{\xi}}}{\textstyle{\kappa_{rr} (\xi) }} \,
                                                 d \xi \, , \\[2mm]
\end{array}
\label{mmnl}
\end{equation}
where $\theta$ is the Heaviside unit step function. The Dirac bracket for
the Lagrangian (\ref{lagpq}) now follows directly from eq.(\ref{defdirac}).

\section{Nonlocal Hamiltonian operators for RHMA$_2$}

   The correspondence (\ref{defhamop}) enables us to express
the result (\ref{mmnl}) for the Dirac bracket of the degenerate Lagrangian
(\ref{lagpq}) in the form of the Hamiltonian operator
\begin{equation}
J_{\kappa} =   \left(            \begin{array}{cc}
0 & \frac{{\textstyle p_x}}{{\textstyle \kappa_{rr} } } D_{x}^{\;-1} \\[3mm]
  D_{x}^{\;-1}  \frac{ {\textstyle p_x}}{{\textstyle \kappa_{rr} } } 
  \;\;\;\;\; &
\frac{{\textstyle q_x}}{{\textstyle \kappa_{rr} }}  D_{x}^{\;-1}  
   + D_{x}^{\;-1} \frac{{\textstyle q_x }}{{\textstyle \kappa_{rr} }}
\end{array}  \right)               \label{jfk}
\end{equation}
where $ D_{x}^{\;-1} $ is the inverse of $ D_{x}$. We refer to
\cite{fokas} for the definition and properties of $ D_{x}^{\;-1} $,
in particular,
$$  D_x^{\;-1} f =  \frac{1}{2}\left( \int^x_{-\infty} -
\int_x^\infty\right) \,  f(\xi) \,  d \xi $$
and the integrals are taken in the principal value sense.
This Hamiltonian operator is therefore nonlocal.
Once again, the appearance of the arbitrary function $\kappa$
in the Hamiltonian operator (\ref{jfk}) is a consequence of the existence
of arbitrary functions in the master Lagrangian (\ref{varpr}).

    The most important member of the family of nonlocal Hamiltonian
operators (\ref{jfk}) is given by
\begin{equation}
J_{1} =   \left(            \begin{array}{cc}
 0 & p_{x}  D_{x}^{\;-1} \\[2mm]
  D_{x}^{\;-1} p_x & q_x  D_{x}^{\;-1} +  D_{x}^{\;-1} q_x
\end{array}  \right)               \label{j2pq}
\end{equation}
which has a linear dependence on $p_{x}$ and $q_{x}$.
It results from the Lagrangian
\begin{equation}
{\cal L}_{1} = \frac{1}{p_x} \left( p_{t} \, q_{x}
 -  \frac{1}{2} q_{x}^{\;2}   \right)
\label{lagrangepq}
\end{equation}
with the simplest choice of arbitrary function
$ \kappa = \frac{1}{2} r^2 $.
It will also be useful to rewrite the operator (\ref{j2pq}) for the system
of variables consisting of $u$ and $q$
\begin{equation}
J_{1} =      \left(            \begin{array}{cc}
 0 & D_{x}^{\;-1} u_{xx}  D_{x}^{\;-1} \\[2mm]
     - D_{x}^{\;-1} u_{xx}  D_{x}^{\;-1}
  & q_x  D_{x}^{\;-1} +  D_{x}^{\;-1} q_x
\end{array}  \right)               \label{j2uq}
\end{equation}
which follows from the change of variable $p = u_x$.

    In section \ref{sec-km} we shall find that the nonlocal Hamiltonian
operator (\ref{j2pq}) has a natural interpretation, namely the Kac-Moody
algebra corresponding to this operator is the algebra of vector fields
and functions on the unit circle. This operator is also distinguished
as the only element of the family of nonlocal Hamiltonian operators
that survives in the case of the real Monge-Amp\`{e}re equation with
constant right hand side.

\subsection{Jacobi identities}

  The proof of the Jacobi identities for $J_{\kappa}$  of eq.(\ref{jfk})
proceeds along the general lines indicated in section \ref{sec-jac23} but
the properties of $D_x^{\;-1}$ must be carefully considered. 
For ease of writing we let $ \rho \equiv (\kappa_{rr})^{-1}$
and from eq.(\ref{jfk}) we have
\begin{equation}
L_{\kappa}  = \left(   \begin{array}{c} 
\rho p_x  D_x^{\;-1} \theta \\[2mm]
D_x^{\;-1} \left( \rho p_x \eta \right) + \rho q_x  D_x^{\;-1} \theta 
+ D_x^{\;-1} \left( \rho q_x \theta \right)
\end{array}  \right)   ,
\label{lj2}
\end{equation}  
$$ I_{\kappa}  = \rho p_x \eta \wedge D_x^{\;-1}   \theta
  + \rho q_x \theta \wedge D_x^{\;-1}   \theta  \, . $$
The variational derivatives of $I_{\kappa}$ are given by
\begin{equation} \begin{array}{lll}
\frac{\textstyle \delta I_{\kappa} }{\textstyle \delta p} & = &
             \left( r \rho_r - \rho \right)
\left( \eta_x \wedge D_x^{\;-1} \theta + \eta \wedge \theta \right)
 + r \rho_{r x} \eta \wedge D_{x}^{\;-1} \theta     \\[2mm]
& & +  \left( r^2 \rho_r \right)_x  \theta \wedge D_x^{\;-1} \theta  
 - r^2 \rho_r \theta_x \wedge D_x^{\;-1} \theta \, ,  \\[3mm]
\frac{\textstyle \delta I_{\kappa} }{\textstyle \delta q} & = &
- \rho_{r x} \eta \wedge D_x^{\;-1} \theta 
- \rho_r \left( \eta_x \wedge D_x^{\;-1} \theta + \eta \wedge \theta \right)
\\[2mm]
& & - \left( \rho + r \rho_r \right) \theta_x \wedge D_x^{\;-1} \theta
- \left( \rho + r \rho_r \right)_x \theta \wedge D_x^{\;-1} \theta
\end{array}
\label{deli2}
\end{equation}
which results in a total derivative for the tri-linear form (\ref{jac})
$$ L_{\kappa}  \delta I_{\kappa}  = - \left\{
  D_x^{\;-1}( p_x \eta + q_x \theta ) \wedge
  \left[ \rho \eta + \left( \rho + r \rho_r \right) \theta \right]
                \wedge D_x^{\;-1}   \theta               \right\}_x $$
completing the proof of the Jacobi identities.

\subsection{Hamilton's equations}

   The Hamiltonian function $H_{\kappa}$ appropriate to the nonlocal
Hamiltonian operator (\ref{jfk}) is given by a local expression which is
simply the integral of the total Hamiltonian density of Dirac in
eq.(\ref{hd}) and once again we have the Lenard-Magri recursion relation
\begin{equation}
 J_{\kappa} \; \delta H_{\kappa} =  J_{\iota} \; \delta H_{\iota}
\label{lem}
\end{equation}
where $\iota$ is another arbitrary function of $p, r$.
All nonlocal Hamiltonian operators are compatible with each other.

\subsection{Kac-Moody algebra}
\label{sec-km}

  Hamiltonian operators associated with integrable nonlinear evolution 
equations give rise to Kac-Moody (KM) algebras \cite{manin}.
Two compatible Hamiltonian operators actually yield an
infinite hierarchy of KM algebras.  
There is an explicit algorithm for the construction of KM algebras from
the Hamiltonian operator which is essentially based on Fourier analysis.
Since the nonlocal operator (\ref{j2pq}) depends linearly
on $p_{x}$ and $q_x$ Fourier analysis makes sense
and the operator (\ref{j2pq}) 
is suitable for consideration as the backbone of a possible KM algebra.
The nonlinearities in the local operator present a
formidable obstacle to any similar discussion of (\ref{jmu}).
Using
$$ p(x) = \frac{1}{2 \pi} \int \frac{1}{n} {\cal P}_n \, e^{i n x} d n \,
\;\;\;\;\;\;\;\;\;
q(x) = \frac{1}{2 \pi} \int \frac{1}{m} {\cal Q}_m \, e^{i m x} d m \, $$
we find that the KM algebra appropriate to the
operator (\ref{j2pq}) is given by
\begin{equation}     \begin{array}{cll}
\left[ {\cal P}_m , {\cal P}_n \right] & = & 0 \; ,\\
\left[ {\cal P}_m , {\cal Q}_n \right] & = &  m \, {\cal P}_{m + n} \; ,   \\
\left[ {\cal Q}_m , {\cal Q}_n \right] & = & ( m - n ) \, {\cal Q}_{m + n}\;,
\end{array}                      \label{km}
\end{equation}
which can be recognized as the algebra of functions and vector fields 
on $S^1$. Thus we can use the representation
$$ {\cal P}_n  =  z^{-n} \, , \hspace{2cm}
 {\cal Q}_m  =  z^{-m+1} \, \frac{d}{d z} $$
for the algebra underlying the simplest  nonlocal Hamiltonian operator
(\ref{j2pq}). 

\subsection{Compatibility}

   Two Hamiltonian operators are compatible if their linear combination
with constant coefficients is also a Hamiltonian operator. In the
case where one of the Hamiltonian operators belongs to the class of
eq.(\ref{jmu}) and the other one to eq.(\ref{jfk}), the check of
compatibility requires that
\begin{equation}
  C_{\mu \kappa}
  = L_{\mu} \, \delta I_{\kappa}  +  L_{\kappa} \, \delta I_{\mu} = 0
\label{comp}
\end{equation}
modulo a divergence. It turns out that only the simplest local and
nonlocal operators $J_0$ and $J_{1}$ are compatible.
From eqs.(\ref{ljmu}), (\ref{delijmu}), (\ref{lj2}) and (\ref{deli2})
it follows that
$$ C_{01} = \left(
 \frac{1}{p_x^{\,2}}\;\theta \wedge \left[
 ( q_x \theta_x + p_x \eta_x ) \wedge D_x^{\;-1} \theta
+ D_x^{\;-1} ( p_x \eta + q_x \theta ) \wedge  \theta_x \right]
\right)_x $$
which establishes the compatibility of $J_0$ and $J_1$.
In all other cases local and nonlocal Hamiltonian operators are
incompatible.

\section{Recursion operator}
\label{sec-rec}

   Since there are infinitely many compatible Hamiltonian operators of both
local (\ref{jmu}) as well as nonlocal (\ref{jfk}) variety, there
exists infinitely many opportunities for constructing recursion operators
of either type or both. In the first category we have
\begin{equation}
 {\cal R}_{\mu \nu} =   J_{\mu} J_{\nu}^{\;-1}   =
   \left(            \begin{array}{cc}
 D_x \frac{\textstyle \nu_q}{\textstyle {\mu_q}} D_{x}^{\;-1} & 0 \\[2mm]
 \left(\frac{\textstyle {\nu_q}}{\textstyle {\mu_q }}  \right)_x
\frac{\textstyle {q_x}}{\textstyle {p_x}} \, D_{x}^{\;-1}  \;\; &
 \frac{\textstyle \nu_q}{\textstyle {\mu_q}}
\end{array}   \right)   \label{calrmu}
\end{equation}
which generalizes our earlier result \cite{ns},
while in the latter category we find
\begin{equation}
{\cal R}_{\kappa \iota} = J_{\kappa} J_{\iota}^{\;-1} =
\left(   \begin{array}{cc}
\frac{\textstyle \iota_{rr}}{\textstyle \kappa_{rr}} & 0 \\[3mm]
D_x^{\;-1} \left[ \frac{\textstyle q_x}{\textstyle p_x}
  \left( \frac{\textstyle \iota_{rr}}{\textstyle \kappa_{rr}} \right)_x
   \right]  &
D_x^{\;-1} \frac{\textstyle \iota_{rr}}{\textstyle \kappa_{rr}} D_x 
\end{array}   \right)             .
\label{camlica}
\end{equation}
These recursion operators satisfy the Lax equation
\begin{equation}
{\cal R}_t = \left[ {\cal R}, {\cal A} \right]
\label{lax}
\end{equation}
where
\begin{equation}
{\cal A} = \left(   \begin{array}{cc}
0 & D_x \\[1mm]
- \frac{\textstyle{q_x^{\;2}}}{\textstyle{p_x^{\;2}}} D_x &
2 \frac{\textstyle q_x}{\textstyle p_x} D_x
\end{array}   \right)
\end{equation}
is obtained from the Frech\'{e}t derivative of the flow (\ref{pq}).

   However, there is also the possibility of constructing recursion operators
by the composition of Hamiltonian operators of both local and nonlocal types
provided they are compatible. We have found that $J_0$ and $J_1$ form
a compatible pair. Thus we are led to the recursion operator
\begin{equation}
 {\cal R} =   J_0 J_{1}^{\;-1} = \left(  \begin{array}{cc}
 D_x \frac{\textstyle 1}{\textstyle p_x} 
        D_x \frac{\textstyle 1}{\textstyle p_x}  & 0 \\[3mm]
 - \frac{\textstyle q_{xx}}{\textstyle{p_x^{\;3}}} D_x -
 D_x  \frac{\textstyle q_{xx}}{\textstyle{p_x^{\;3}}} \;\; &
 \frac{\textstyle 1}{\textstyle p_x} D_x
    \frac{\textstyle 1}{\textstyle p_x} D_x  \end{array}  \right)
\label{re2}
\end{equation}
and we find that a remarkable factorization takes place
\begin{equation}
 {\cal R} =  {\cal E}  {\cal E}
\label{re2ee}
\end{equation}
where 
\begin{equation}
 {\cal E} =    \left(            \begin{array}{cc}
  D_{x} \frac{{\textstyle 1}}{{\textstyle p_x}} & 0 \\[2mm]
 -  \frac{{\textstyle q_{xx}}}{{\textstyle p_x^{\;2}}} &
 \frac{{\textstyle 1}}{{\textstyle p_x}}   D_{x}
\end{array}   \right)                 \label{calef}
\end{equation}
is a first order operator.
This situation is familiar from third order Hamiltonian operators for two
component equations of hydrodynamic type \cite{sh}. The recursion operator
obtained from a composition of third and first order Hamiltonian operators
factorizes to yield the first order recursion operator of Sheftel'. Similar
considerations suggest that ${\cal E}$ itself is a good recursion operator.
It is readily verified that ${\cal E}$ as well as its inverse
\begin{equation}
 {\cal E}^{-1} =    \left(            \begin{array}{cc}
 p_x D_{x}^{\;-1} & 0 \\[1mm]
 q_x D_{x}^{\;-1} - D_{x}^{\;-1}     q_x &  D_{x}^{\;-1} p_x
\end{array}   \right)   \label{cale}
\end{equation}
satisfy eq.(\ref{lax}) and are therefore good recursion operators.
Finally, we should remark that the recursion operators (\ref{calrmu}) and
(\ref{camlica}) can also be factorized, however, unlike the case of
eq.(\ref{re2ee}), the factors consist of different operators and each
one is not a recursion operator.

\subsection{Infinite sequences of Hamiltonians}

    There are several infinite families of conserved quantities
that one can obtain by repeated application of the recursion operators
we have presented in section \ref{sec-rec}.
Corresponding to the recursion operator $ {\cal E} $ we have ${\cal F}$
which is defined by
\begin{equation}
  {\cal S} = J_{0}^{\;\;-1} J_1 = {\cal F}  {\cal F}
\label{k0j2}
\end{equation}
which yields infinite sequences of gradients of conserved quantities.
We find that $ {\cal F} $ is given by
\begin{equation}
 {\cal F} =    \left(            \begin{array}{cc}
 D_{x}^{\;-1}  p_x &  D_{x}^{\;-1} q_x - q_x  D_{x}^{\;-1} \\[1mm]
 0 & p_x  D_{x}^{\;-1} 
\end{array}   \right)   \label{fk0j2}
\end{equation}
and a simple example of the infinite sequence of conserved quantities
that it generates is the following
\begin{equation}
    \begin{array}{ccccc}
    &  \vspace{-0.4cm}  {\cal F} &     & {\cal F}        &      \\
 p \, q_x & \longrightarrow         & ... & \longrightarrow &
 \frac{\textstyle 1}{\textstyle n!} \, p^n \, q_x  \, .
\end{array}                      \label{conse}
\end{equation}
Depending on the starting point one can construct infinitely
many such sequences.
For example, ${\cal S}$ generates a family of local as well as
nonlocal conserved quantities according to the scheme
\begin{equation}
    \begin{array}{ccccccc}
    &  \vspace{-0.4cm} {\cal S}   &   & {\cal S}     & &  &      \\
... & \longrightarrow &
\frac{\textstyle 1}{\textstyle 2} \, q D_x \left\{
\frac{\textstyle 1}{\textstyle p_x} D_x \left[
\frac{\textstyle 1}{\textstyle p_x} D_x \left(
\frac{\textstyle q_x}{\textstyle p_x} \right)  \right]  \right\}
& \longrightarrow &
 -\frac{\textstyle 1}{\textstyle 2}  \,
\frac{\textstyle {q_{x}^{\;2}}}{\textstyle p_x} & \\[5mm]
 &  \vspace{-0.3cm}  {\cal S}       &  \hspace{-4cm}     &
  \hspace{-8cm} {\cal S} & \hspace{-5cm} & \hspace{-1cm} {\cal S} &   \\
 & \longrightarrow &
\hspace{-4cm} \frac{\textstyle 1}{\textstyle 2} \, q^2 \, p_x
& \hspace{-8cm} \longrightarrow & \hspace{-5cm}
\frac{\textstyle 1}{\textstyle 2} \, q \, p_x D_{x}^{\;-1} \left[
p_x D_{x}^{\;-1} \left( q p_x \right) \right]
& \hspace{-1cm} \longrightarrow & ... \, .
\end{array}                      \label{consr}
\end{equation}
and there are infinitely many different combinations of such sequences. The
above expressions for the conserved Hamiltonian densities can be simplified
by discarding divergences, however, this form is useful because it suggests
the generic term for the conserved quantity obtained by repeated applications
of ${\cal S}$. The determination of the cardinality of the conserved
Hamiltonians which are in involution with respect to Poisson brackets defined
by both types of Hamiltonian operators is complicated because of the
appearance of arbitrary functions in these Hamiltonian operators.
      
\subsection{Higher flows}
     
     Starting with the flow (\ref{pq}) we should obtain higher flows for
RHMA$_2$ through the application of any one of the recursion operators
(\ref{calrmu}), (\ref{camlica}), (\ref{calef}), or (\ref{cale})
on the vector field (\ref{pq}). However, in all of these cases an
elementary calculation shows that the resulting higher flow is simply
$$ p_{t} \, q_{x} - q_{t} \, p_{x} = 0 \, , $$
the real homogeneous Monge-Amp\`{e}re equation back again.

\section{Symplectic form of RHMA$_2$}
\label{sec-symp}

   The principal geometrical object in the theory of symplectic structure
is the symplectic 2-form $\omega$ which is closed
\begin{equation}
d \omega  = 0                      \, ,
\label{closed}
\end{equation}
and by Poincar\'e's lemma $\omega$ can be written as
\begin{equation}
\omega =  d \alpha
\label{alphao}
\end{equation}
in a local neighborhood. On the other hand the Hamiltonian operator maps
differentials of functions into vector fields which works in the opposite
direction. Thus the statement of the symplectic structure of the equations
of motion consists of
\begin{equation}
 i_X \omega  =  d H   \label{symhameq}
\end{equation}
which is obtained by the contraction of the symplectic 2-form $\omega$ 
with the vector field {\bf X} defining the flow.

    For systems with finite, and furthermore even number of degrees of
freedom, the symplectic 2-form is the inverse of the Hamiltonian
structure functions which is the analog of the Hamiltonian operator.
However, the generalization of the notion of an inverse to systems with
infinitely many degrees of freedom is not immediate. That is,
the symplectic 2-form is obtained by integrating the density \cite{dm}
\begin{equation}
\omega = \frac{1}{2} \, d u^i \wedge K_{ij} \, d u^j
\label{defomega}
\end{equation}
over the spatial variable, where $K$ is the ``inverse" of $J$.
Given Hamiltonian operator $J$, its ``inverse" may be defined by
\begin{equation}
 J^{i k} K_{k j} = 
 K_{j k} J^{k i} = \delta^{i}_{j}
\label{invjk}
\end{equation}
but eq.(\ref{invjk}) is an operator equation which acts on
gradients of functions.
On the other hand, quite generally there exist Casimir functions which
are annihilated by the Hamiltonian operator. Such functions, {\it c.f.}
eq.(\ref{casimir}) above, must be excluded in the definition of the
``inverse" in eq.(\ref{invjk}).

    Our approach to the construction of Hamiltonian operators is based on
the construction of Dirac brackets for systems subject to
second class primary constraints.
It is evident from eq.(\ref{defdirac}) that the essential element in
the Dirac bracket is the inverse of the matrix of Poisson brackets of
the constraints. On the other hand, according to eq.(\ref{invjk})
we must invert again to obtain the symplectic 2-form, thus we have simply
\begin{equation}
 K_{ik} (x)  \delta ( x - y )  \equiv \{ \phi_i (x) , \phi_k (y) \} ,
\label{pbceqinv}
\end{equation}
that is, the inverse of the Hamiltonian operator can be obtained directly
from the Poisson bracket of second class constraints.

\subsection{Symplectic 2-forms}

  The Hamiltonian operator (\ref{jmu}) can be inverted in a straight-forward
way subject to the provision above
\begin{equation}
 K_{\mu} =    \left(            \begin{array}{cc}
-     \frac{ {\textstyle \mu_q q_x}}{\textstyle q} D_{x}^{\;-1}
-  D_{x}^{\;-1}   \frac{ {\textstyle \mu_q q_x}}{\textstyle q} &
D_{x}^{\;-1}     \frac{ {\textstyle \mu_q p_x}}{\textstyle q} \\[2mm]
     \frac{ {\textstyle \mu_q p_x}}{\textstyle q}  D_{x}^{\;-1} & 0
\end{array}   \right) ,  \label{kmu}
\end{equation}
or we can immediately read it off from eqs.(\ref{pbconst}) and
(\ref{pbceqinv}). Eq.(\ref{kmu}) is also a statement of the non-degeneracy
of the Hamiltonian operator (\ref{jmu}). Hence from eqs.(\ref{defomega})
and (\ref{kmu}) we find the symplectic 2-form
\begin{equation}
\omega_{\mu}  =     \frac{ {\textstyle \mu_q}}{{\textstyle q}}
     \left( p_{x} \, d q - q_x \, d p \right) \wedge d (D_x^{\;-1}p)
\label{omu}
\end{equation}
corresponding to the local Hamiltonian operator (\ref{jmu}).
It can be verified that $\omega_{\mu}$ is a closed 2-form by
direct calculation.
We had found \cite{ns} that in the $u, q$ variables the
inverse of $J_0$ is a local operator and this is true
for the family of local Hamiltonian operators (\ref{jmu}) in general.
The symplectic 2-form assumes the simpler expression
\begin{equation}
\omega_{\mu}  = \lambda_{u_{x} u_{x}}
     \left( u_{xx} \, d q - q_x \, d u_{x}  \right) \wedge d u
\label{omuq}
\end{equation}
in these variables.
By invoking the Poincar\'e lemma, in a local neighborhood we can write
\begin{equation}   
\omega_{\mu}  =  d \alpha_{\mu} \, \;\;\;\;\;\;\;\;\;
\alpha_{\mu}  =  \sigma p_x \, d (D_x^{\;-1}p)  
\label{amu}
\end{equation}
and we note that the coefficient of the 1-form $\alpha_{\mu}$ is also the
Casimir (\ref{casimir}) for the Hamiltonian operator $J_{\mu}$.
The closure of the symplectic 2-form (\ref{omu}) is equivalent to the
satisfaction of the Jacobi identities by the Hamiltonian
operator (\ref{jmu}).

   For the nonlocal Hamiltonian operator (\ref{jfk}) the inverse
is given by the local operator
\begin{equation}
 K_{\kappa} =    \left(            \begin{array}{cc}
-   \frac{ {\textstyle q_x \kappa_{rr} }}{\textstyle p_{x}^{\;2}} D_{x}
-  D_{x} \frac{ {\textstyle q_x \kappa_{rr} }}{\textstyle p_{x}^{\;2}} &
     \frac{ {\textstyle \kappa_{rr} }}{\textstyle p_x}  D_x  \\[1mm]
 D_x     \frac{ {\textstyle \kappa_{rr} }}{\textstyle p_x}  & 0
\end{array}   \right)   \label{k2}
\end{equation}
which also follows from eqs.(\ref{pbconstnl}) and (\ref{pbceqinv}).
Then from eqs.(\ref{defomega}) and (\ref{k2}) the symplectic 2-form
appropriate to the nonlocal Hamiltonian operator is given by
\begin{equation} \begin{array}{rcl}
\omega_{\kappa}  & = & \kappa_{rr}  \,
\left( - \frac{\textstyle q_x}{\textstyle{p_x^{\;2}}} d  p_x \,
 + \frac{\textstyle 1}{\textstyle p_x} \,  d q_x \right) \wedge d p \\[4mm]
 & = & \kappa_{rr}  \,
d r \wedge d p
\end{array}
\label{o2}
\end{equation}
so that the verification that $\omega_{\kappa}$ is closed is immediate.
In a local neighborhood we can write
\begin{equation}  
\omega_{\kappa}   = d \alpha_{\kappa} ,    \;\;\;\;\;\;
\alpha_{\kappa}   =  \kappa_r \, d p 
\label{a2}
\end{equation}
using the Poincar\'e lemma.

\subsection{Symplectic form of equations of motion}

  With the symplectic 2-forms (\ref{omu}) and (\ref{o2}) we need to
check that eqs.(\ref{symhameq}) are satisfied.
For this purpose we recall that given a 2-form $ \omega = a(v, v_x)
 \, d v \wedge d v_x $ and the vector field
$ X = m(v, v_x) \, \partial / \partial v $, we have
$ i_X \omega = ( 2 a m_x + m a_x ) \, d v $.
If we consider $\omega_{\kappa}$ with
$\kappa = \frac{1}{2} r^2$, from this expression we get
\begin{equation} \begin{array}{lll}
i_X \, \omega_{\frac{1}{2} r^2}
 &= & - \frac{{\textstyle q_x}}{{\textstyle p_x}} d q_x
-  i_{ \left( q_x \frac{\partial}{\partial p} \right) }
\frac{{\textstyle q_x}}{{\textstyle p_x^{\;2}}} dp \wedge  d p_x  
+ i_{ \left( \frac{q_{x}^{\,2}}{p_x} \frac{\partial}{\partial q} \right) } 
 \frac{{\textstyle 1}}{{\textstyle p_x}}  dp \wedge d q_x  \\[4mm]
&=& - \frac{{\textstyle q_x}}{{\textstyle p_x}} d q_x
-\left[ 2 \frac{{\textstyle q_x }}{{\textstyle p_x^{\;2}}} q_{xx} +
q_x \left( \frac{{\textstyle q_x }}{{\textstyle p_x^{\;2}}} \right)_x
- \frac{{\textstyle q_x^{\;2} }}{{\textstyle p_x^{\;3}}} p_{xx} \right] dp
- \frac{{\textstyle q_x^{\;2} }}{{\textstyle p_x^{\;2}}} d p_x \\[4mm]
&=& d \left(
- \frac{{\textstyle 1}}{{\textstyle 2}}
\frac{{\textstyle q_x^{\;2} }}{{\textstyle p_x}} \right)  \;
= \; d H_{\frac{1}{2} r^2}
\end{array}  \label{eqsympl}
\end{equation}
where we have discarded a total derivative.
Similarly, in order to check that
$ i_X  \, \omega_{\mu}  = d H_{\mu} $,
we note that given the 2-form $ \omega = a(v, v_x, ...)
 \, d v \wedge d (D_x^{\;-1} v ) $ and the vector field
$ X = m(v, v_x, ...) \, \partial / \partial v $,
we have
$ i_X \omega  = - a m d ( D_x^{\;-1} v ) - m  D_x^{\;-1} ( a d v ) \,.$
The application to the 2-form (\ref{omu}) yields
\begin{equation}           \begin{array}{lll}
i_X \, \omega_{\mu} & = & - q_x D_x^{\;-1} ( \sigma_q p_x dq )
+ q_x D_x^{\;-1} ( \sigma_q q_x dp ) \\[1mm]
& = & q \sigma_q ( p_x dq - q_x dp ) \; = \; d H_{\mu}
\end{array}
\end{equation}
using eq.(\ref{defsig}).

\subsection{Witten-Zuckerman 2-form}

      Time plays a privileged role in Hamiltonian mechanics.
While this presents no problem for systems with finitely many degrees
of freedom, in field theory it has the disadvantage of non-covariance.
In order to remedy this situation Crnkovi\'{c} and Witten and
Zuckerman \cite{cwz} have introduced the conserved current 2-form
which provides an elegant covariant formulation of Hamiltonian structure.
For Monge-Amp\`{e}re covariance is particularly necessary
because, as we noted in the introduction, the choice of time coordinate
for RHMA$_2$ is quite arbitrary.

   The simplest way to obtain the Witten-Zuckerman current 2-form $\omega$
for the Lagrangian (\ref{lmu}) is to first construct the 1-form $\alpha$
which follows from the first variation of the Lagrangian and $\alpha$ is
related to $\omega$ as in eq.(\ref{alphao}). Assuming the equations
of motion (\ref{uq}), the first variation reduces to a conservation law
\begin{equation}
\delta {\cal L}_{\lambda}  = \alpha^{t}_{\;\lambda, \;t}
+ \alpha^{x}_{\;\lambda, \; x} \, ,
\end{equation}
where
\begin{equation}    \begin{array}{lll}
\alpha^{t}_{\;\lambda} & = &   \frac{\textstyle {\partial {\cal L}_{\lambda}
    } }   {\textstyle {\partial u_t} }  \delta u
 + \frac{\textstyle {\partial  {\cal L}_{\lambda}  } }
 {\textstyle {\partial q_t} } \delta q =
 - q_{x} \, \lambda_{u_{x}} \, \delta u  +  \lambda \, \delta q  \, , \\[4mm]
\alpha^{x}_{\;\lambda} & = &   \frac{\textstyle {\partial {\cal L}_{\lambda}
  } }               {\textstyle {\partial u_x} }  \delta u
 + \frac{\textstyle {\partial  {\cal L}_{\lambda}  } }
 {\textstyle {\partial q_x} } \delta q
 =  \lambda_{u_{x}} \, q_{t} \, \delta u .
\end{array}
\label{varfirst}
\end{equation}
Finally, the current 2-form $\omega_{\lambda}$ for RHMA$_2$ is given by
\begin{equation}    \begin{array}{lll}
\omega^{t}_{\;\lambda} & = & \delta \alpha^{t}
 =  \lambda_{u_{x} u_{x}} \left( q_{x} \, \delta u \wedge \delta u_{x}
 - u_{xx} \, \delta u \wedge \delta q     \right)   \\[4mm]
\omega^{x}_{\;\lambda}  & = & \delta \alpha^{x} =
 \lambda_{u_{x} u_{x}}    \left( -
 \frac{\textstyle { q_{x}^{\,2} }}{\textstyle { u_{xx} } }  \,
 \delta u \wedge \delta u_{x} \,
 + \, q_{x} \, \delta u \wedge \delta q \right)
\end{array}
\label{wz2fl}
\end{equation}
and using the equations of motion we can readily verify that
it satisfies
\begin{equation}
\delta \omega^{\alpha} = 0, \hspace{1cm}
\omega^{\alpha}_{\;,\alpha} = 0,
\label{wz2fprop}
\end{equation}
where $\alpha$ ranges over two values $t$ and $x$.
The Witten-Zuckerman 2-form is closed and conserved.
For the Lagrangian (\ref{lagpq}) a similar procedure yields
\begin{equation}    \begin{array}{lll}
\omega^{t}_{\;\kappa} & = &  \kappa_{rr} \, \delta r \wedge \delta p  \\[3mm]
\omega^{x}_{\;\kappa} & = & r \,  \kappa_{rr} \, \delta p \wedge \delta r
\end{array}
\label{wz2fnl}
\end{equation}
which also satisfies eqs.(\ref{wz2fprop}). A useful relation
in checking the conservation law for the $2$-form (\ref{wz2fnl})
is the dKdV, or the Riemann equation form of RHMA$_2$
\begin{equation}
r_{t} = r \, r_{x}
\end{equation}
which follows from the results of \cite{jmn}.

   We note that the time components of the Witten-Zuckerman 2-forms
(\ref{wz2fl}) and (\ref{wz2fnl}) for RHMA$_2$ that follow from the
Lagrangians (\ref{lmu}) and (\ref{lagpq}) are precisely the symplectic
2-forms (\ref{omuq}) and (\ref{o2}) respectively.
The use of the notation $\delta$ for $d$ follows \cite{cwz} and is
restricted to this section only.

\section{Lax Pair for RHMA$_2$}

   We have seen that the recursion operators of section \ref{sec-rec}
satisfy the Lax equation (\ref{lax}) but these are not useful Lax pairs.
We need to cast RHMA$_2$ into the form of a zero-curvature condition
\cite{akns}
\begin{equation}
U_t - V_x - \left[ U , V \right] = 0
\label{laxeq}
\end{equation}
which is the basic element in the solution of of completely integrable
systems using the inverse scattering transform. The zero-curvature condition
for the $SL(2,R)$-valued connection 1-form given by the pair
\begin{equation}
 U =  \left(   \begin{array}{cc}       \lambda & p_x \\[2mm]
 q \, p_x - \frac{\textstyle \lambda}{\textstyle p_{x}^{\,2}} p_{xx}
          - \frac{\textstyle \lambda^2}{\textstyle p_{x}}
 & - \lambda    \end{array}   \right) , \;\;
 V = \left(  \begin{array}{cc}        \lambda
 \frac{\textstyle q_x}{\textstyle  p_x} & q_x \\
 q \, q_x - \frac{\textstyle \lambda}{\textstyle p_{x}^{\,2}} q_{xx}
          - \frac{\textstyle \lambda^2}{\textstyle p_{x}^{\,2}} q_x
 & -     \lambda  \frac{\textstyle q_x}{\textstyle  p_x}
   \end{array}   \right)
\label{laxpair}
\end{equation}
provides such a formulation of RHMA$_2$.
But in this case the potential has a quadratic dependence on the
spectral parameter $\lambda$ which has so far not been considered
for an application of inverse scattering techniques.
This $U, V$ pair is therefore not immediately amenable to
treatment by the method of inverse scattering.

\section{Multi-Hamiltonian structure of RMA$_2$}

    The infinite classes of Hamiltonian operators we have obtained
for RHMA$_2$ reduce to the compatible pair $J_0$ and $J_1$ of
Hamiltonian operators (\ref{jmu01}) and (\ref{j2pq}) when we consider the
Hamiltonian structure of RMA$_2$ with non-zero constant right hand side in
eq.(\ref{rhma2}). The corresponding pair of symplectic 2-forms are also
unchanged and the only modification comes in the conserved quantities.
The infinite sequence of conserved Hamiltonians for RMA$_2$ are those
which reduce to the RHMA$_2$ Hamiltonians for $ \mu = \frac{1}{2} q^2 $ and
$ \kappa = \frac{1}{2} r^2 $ in the limit $ K \rightarrow 0$.
Thus the basic Hamiltonian densities entering into the Lenard-Magri scheme
\begin{equation}
 J_0 \, \delta \, H_{1}^{K} = J_1 \, \delta \, H_{0}^{K}
\label{hamrma}
\end{equation}
are given by
\begin{equation}
{\cal H}_{1}^{K} = \frac{1}{2} q^2 p_x + K \, D_{x}^{\;\;-1} p \, ,
\hspace{1cm}
{\cal H}_{0}^{K} = - \frac{1}{2 \, p_x} \left( q_{x}^{\;\;2} + K \right)
\label{hrma}
\end{equation}
and the infinite sequences of conserved
Hamiltonians which are in involution with respect to Poisson
brackets defined by $J_0$ and $J_1$ are modified for
RMA$_2$. For example, by the application of the recursion operator
(\ref{calef}) to $H_{1}^{K}$ we get
\begin{equation}
{\cal H}_{2}^{K} = \frac{\textstyle 1}{\textstyle 2 p_{x}}
  \left[ \left(
\frac{\textstyle q_x}{\textstyle p_x} \right)_{\!x}  \right]^2
 + \frac{\textstyle K}{\textstyle 2}
\frac{\textstyle{ p_{xx}^{\;\;2} }}{\textstyle{ p_{x}^{\;5} }}
\end{equation}
which, up to a divergence, is the same as the RHMA$_2$
Hamiltonian density in the sequence (\ref{consr}) for $K \rightarrow 0$.
Repeated application of the recursion operator (\ref{calef}) yields
\begin{equation}               \begin{array}{rcl}
p_{t} \, q_{x} - q_{t} \, p_{x} & = & -
 D_x \left\{
\frac{\textstyle 1}{\textstyle p_x} D_x \left[
\frac{\textstyle 1}{\textstyle p_x}   \;...\; K \;...\;
 p_{x} D_{x}^{\;\,-1} \left( p_{x} D_{x}^{\;\,-1}
  \right)  \right]  \right\}
\end{array}
\label{sirma}
\end{equation}
for the RMA$_2$ hierarchy of equations.

   The elliptic case of eq.(\ref{rhma2}) is equivalent to the equation for
minimal surfaces while the hyperbolic case corresponds to the Born-Infeld
equation \cite{jmn}. Through the appropriate change of variables the rich
multi-Hamiltonian structure of the Born-Infeld equation \cite{annov}
carries over into RMA$_2$ which includes the Hamiltonian operators $J_0$
and $J_1$.

\section{Ur-RHMA$_2$}
\label{sec-urma2}

    The local and nonlocal family of Hamiltonian operators for RHMA$_2$
have scalar counterparts for the Ur-RHMA$_2$ equation (\ref{urma}) which
can be written as
\begin{equation}
u_t  = \frac{k}{u_x}
\label{urmaev}
\end{equation}
in  the form of an evolution equation. With the definition
$$ U  =  \frac{2 k}{u_{x}^{\;2}} $$
this equation can be identified as
$$ U_t  + U \, U_x = 0 $$
which is the dispersionless KdV, or Riemann equation. Ur-RHMA$_2$ admits
infinitely many conserved quantities
\begin{equation}
{\cal H}_n  = u_{x}^{\;n}
\label{hurma}
\end{equation}
as well as infinitely many local Hamiltonian operators
\begin{equation}
J_{\alpha}  = \frac{k_{\alpha}}{ u_{x}^{\;\alpha} u_{xx} } D_x
\frac{1}{ u_{x}^{\;\alpha} u_{xx} }
\label{jurmaloc}
\end{equation}
provided the various constants are related by
$$ n = 2 ( \alpha + 1 ), \;\;\;\;\;
2 \alpha ( \alpha + 1 ) ( 2 \alpha + 1 ) k_{\alpha}  = - k $$
so that the equation of motion (\ref{urmaev}) assumes the form of
Hamilton's equations (\ref{urhameq}).
Scalar Hamiltonian operators of this type were first considered by Vinogradov
\cite{vm}. The family of nonlocal Hamiltonian operators
\begin{equation}
J_{\beta}  = k_{\beta} u_{x}^{\;\beta}  D_{x}^{\;-1} u_{x}^{\;\beta}
\label{jurmanonloc}
\end{equation}
which is due to Sokolov \cite{so} is also appropriate to
the Ur-RHMA$_2$ equation. In this case the conserved Hamiltonian densities
are also given by eq.(\ref{hurma}) but now the constants are related by
$$ n = - 2 \beta , \;\;\;\;\;
  2 \beta ( 2 \beta + 1 ) k_{\beta}  = ( \beta + 1 ) k $$
and we find that Ur-RHMA$_2$ is cast into Hamiltonian form with
\begin{equation}
u_{t} = J_{\alpha} \, \delta H_{2 \alpha + 2 } \;
                  = \; J_{\beta} \, \delta H_{-2\beta}
\label{urhameq}
\end{equation}
which again results in an infinite set of Hamiltonian structures.
In eqs.(\ref{jurmaloc}) and (\ref{jurmanonloc}) we have the $qq$-components
of the RHMA$_2$ matrix Hamiltonian operators (\ref{jmu}) and
(\ref{jfk}) respectively.

    The recursion operator obtained by the composition of these
Hamiltonian operators
\begin{equation}
{\cal R}_{\alpha \beta}  = \frac{1}{u_{x}^{\;\alpha} u_{xx} }  D_{x}
 \frac{1}{u_{x}^{\;\alpha + \beta} u_{xx} }  D_{x} \frac{1}{u_{x}^{\beta}}
\label{rurma}
\end{equation}
can be factored as in eq.(\ref{re2ee}) with
\begin{equation}
{\cal E}_{\alpha \beta}  = \frac{1}{u_{x}^{\;\alpha} \, u_{xx} }  D_{x}
 \frac{1}{u_{x}^{\;\beta}}
\label{eurma}
\end{equation}
resulting in a Sheftel'-type recursion operator for Ur-RHMA$_2$.

\section{RHMA in arbitrary dimension}
\label{sec-rhman}

    In order to write RHMA$_n$ as a system of nonlinear evolution equations
it will be useful to introduce a compact notation. For this purpose
we shall consider the determinants of the $n-1 \times n-1$ matrices
\begin{equation}
  \Delta^{k} \equiv      (- 1)^{k+1}       \det
\left(       \begin{array}{cccccc}
q_1 & u_{1\,1} & ... & \widehat{u_{1\,k}} & ... & u_{1\,{n-1}} \\
q_2 & u_{2\,1} & ... & \widehat{u_{2\,k}} & ... & u_{2\,{n-1}} \\
... & ...      &     & ...                &     & ... \\
q_{n-1} & u_{{n-1}\,1} & ... & \widehat{u_{n-1\,k}} & ... & u_{n-1\,n-1}
\end{array}   \right)
\label{cofactor}
\end{equation}
obtained by deleting the $0^{th}$ row and $k^{th}$ column
in the matrix of second derivatives.
The latter is indicated by a hat
over the omitted terms.
In particular, for $k = 0$ we have
the Monge-Amp\`ere operator in $n-1$ dimensions
$$   \Delta \equiv - \Delta^{0} \ne 0   $$
which is a statement of nondegeneracy of RHMA$_n$.
The system of evolution equations for RHMA$_n$ is given by
\begin{equation}   \begin{array}{ccl}
u_t &=& q \; , \\[2mm]
q_t &=& \frac{{\textstyle 1}}{{\textstyle \Delta}} q_i \Delta^i
\; ,    \;\;\;\;\;\;\;\;\;\; i=1,2,...,n-1
\end{array}    \label{evn}
\end{equation}
where $ q_i = \partial q / \partial x^i $ and
henceforth we shall reserve the index $i$ to range over $n-1$
independent variables while continuing the use of the summation
convention over repeated indices. Thus the vector field
\begin{equation}
{\bf X} = q \frac{\partial}{\partial u}
+ \frac{{\textstyle 1}}{{\textstyle \Delta}} q_i \Delta^i
  \frac{\partial}{\partial q}
\label{evnx}
\end{equation}
defines the flow for RHMA in $n$ dimensions.

   Eqs.(\ref{evn}) are cast into the form of Hamilton's equations
with the Hamiltonian operator
\begin{equation}
{\cal J} =   \left(            \begin{array}{cc}
 0 & \frac{ {\textstyle  q } }{  {\textstyle \mu' \Delta } } \\[4mm]
 -  \frac{ {\textstyle  q } }{  {\textstyle \mu' \Delta } } \;\;\;
&   \frac{ {\textstyle  q \Delta^i} }{  {\textstyle {\mu' \Delta^2} } }
D_i + D_i
 \frac{ {\textstyle  q \Delta^i} }{  {\textstyle {\mu' \Delta^2} } }
\end{array}  \right)               \label{jdelta}
\end{equation}
where $\mu$ is an arbitrary differentiable function of $q$ alone 
and prime denotes derivative with respect to the argument.
The Hamiltonian function is given by
\begin{equation}
{\cal H} = \mu \Delta               \label{hn}
\end{equation}
and once again, there exist infinitely
many conserved quantities (\ref{hn}) and Hamiltonian operators
(\ref{jdelta}) associated with RHMA in $n$ dimensions.
However, this is not the full extent of the Hamiltonian structure
of eqs.(\ref{evn}) as we have not considered nonlocal operators,
or the dependence of $\mu$ on other variables. Concerning the latter 
point we note that
\begin{equation}
{\widetilde {\cal H}} = \mu(q,u_1,u_2,...,u_{n-1}) \Delta
\label{hnext}
\end{equation}
is also conserved for the system (\ref{evn}).
Hence the number of independent variables entering into the arbitrary 
function $\mu$ can be increased considerably
with an attendant increase in the number of Hamiltonian operators
which is already infinite in eq.(\ref{jdelta}).

  In order to present the symplectic structure of eqs.(\ref{evn})
we need the inverse of (\ref{jdelta}) and
subject to the provisions of section \ref{sec-symp}, we find
that it is again a local operator. Then from eq.(\ref{defomega}) we
get the symplectic 2-form
\begin{equation}
\omega = \frac{{\textstyle \mu'}}{{\textstyle {q}}} \left(
\Delta^i du \wedge du_i  + \Delta dq \wedge du   \right)
\end{equation}
which can be directly verified to be a closed 2-form.
In a local neighborhood we can write it as the exterior derivative of
a 1-form $\alpha$ which is given by
\begin{equation}
\alpha = \sigma \,\Delta \, du
\end{equation}
where $\sigma$ is again related to $\mu$ through eq.(\ref{defsig}) and
$\sigma \Delta$ is the Casimir for the Hamiltonian operator (\ref{jdelta}).

\section{Geodesic flow for CHMA}
\label{sec-gchma}

   The Hamiltonian structure of the geodesic flow for CHMA is very similar
to that of RHMA. Semmes \cite{semmes} has introduced the notion of
geodesics on ${\cal N}$, the space of smooth real-valued functions on
 $ I \times M $ where $I$ is a real interval.
For $ F \in {\cal N} ( I \times  M ) $ and
$ ( \partial {\bar {\partial} } F )^n  \ne 0 $ the vector field
\begin{equation}
X = q \frac{\partial}{\partial F}  + n
\frac{  \left[      ( \partial {\bar {\partial} } F )^{n - 1} \wedge
 \partial q \wedge {\bar {\partial} } q  \right]   }
{ \left[ ( \partial {\bar {\partial} } F )^n \right] }
 \,  \frac{\partial}{\partial q}
\label{geodesic}
\end{equation}
defines geodesics on  ${\cal N}$. The holomorphic exterior derivative
is denoted by $\partial$.
Here as well as in the following, it will be understood that
volume forms on $M$ enclosed by square parantheses automatically carry
the Hodge star operator so that the result is a 0-form.
The discussion of the symplectic structure of CHMA by Semmes is
based on the K\"{a}hler 2-form
$$   \frac{1}{2i}  \partial {\bar {\partial} } F  $$
which is not the relevant object that emerges from an examination of
the Hamiltonian structure of the flow (\ref{geodesic}).
Our approach to the problem of the geodesic flow for CHMA
will be in the framework of dynamical systems with infinitely
many degrees of freedom and the resulting symplectic 2-form
is given in eq.(\ref{symgchma}).
The advantage of our approach lies in the direct proof it furnishes for
the complete integrability of the geodesic flow for CHMA.

   The geodesic flow for CHMA satisfies Hamilton's equations
\begin{equation}
\left( \begin{array}{l} F_t \\ q_t \end{array} \right) =
{\bf X} \left( \begin{array}{l} F \\ q \end{array} \right) =
        {\cal J}_{c} \delta H_{c}
\label{hameqg}
\end{equation}
where ${\bf X}$ denotes the vector field (\ref{geodesic}).
The Hamiltonian density is given by
\begin{equation}
{\cal H}_{c} = \mu \left[  ( \partial {\bar {\partial} } F )^n \right]
\label{hgchma}
\end{equation}
and the Hamiltonian operator is
\begin{equation}
{\cal J}_{c} =
\left(            \begin{array}{cc}
 0 & \frac{ {\textstyle  q} }{  {\textstyle \mu'
     \left[  ( \partial {\bar {\partial} } F )^n \right] } } \\[4mm]
 \frac{ {\textstyle - q}}{  {\textstyle \mu'
 \left[ ( \partial {\bar {\partial} } F )^n \right] } } &
          Re    \left\{
 \frac{ {\textstyle n q \left[ {\bar \partial q}  \wedge
      (   \partial {\bar {\partial} }  F )^{n - 1} \right. } }
{{\textstyle \mu' \,\left[ (\partial {\bar {\partial}} F )^{n} \right]^2 }  }
          \wedge  \partial \left. \right]
-   \left[ \partial  \wedge \right.
\frac{ {\textstyle n q  {\bar \partial} q  \wedge
    ( \partial {\bar {\partial} } F )^{n - 1}
             \left. \right]        } }
{{\textstyle \mu' \,\left[ (\partial {\bar {\partial}} F )^{n} \right]^2 }  }
 \right\}
\end{array}  \right)  
\label{jgchma}
\end{equation}
where $\mu=\mu(q)$ is an arbitrary differentiable function of its argument.

  The inverse of Hamiltonian operator (\ref{jgchma}), subject to the 
restrictions of section \ref{sec-symp}, is again a local operator
which yields the symplectic 2-form
\begin{equation}         \begin{array}{l}
\omega_c = \frac{\textstyle{\mu'}}{\textstyle q}
              \left\{  \right.  Re  \left( d F \wedge
\left[   \partial q  \wedge
( \partial {\bar {\partial} } F )^{n-1} \wedge
    {\bar   \partial } \right] d F \right)  \, + \,
     \frac{ {\textstyle 1}}{{\textstyle n}}
     \left[ ( \partial {\bar {\partial} } F )^n \right]
d q \wedge d F \left. \right\} 
\end{array}
\label{symgchma}
\end{equation}
and for integrable complex structure $\omega$ can be simplified by
expressing the exterior derivative in terms of
$  \partial, {\bar {\partial} } $.
The statement of the symplectic structure of
the geodesic flow for CHMA is given by eq.(\ref{symhameq}).
The 2-form (\ref{symgchma}) is closed
as one can show readily by direct calculation.
However, it is more instructive to note that by invoking the
Poincar\'e lemma in a local neighborhood we can write
\begin{equation}                    
\omega_c =  d \alpha_c \, , \;\;\;\;\;
\alpha_c =   \frac{{\textstyle 1}}{{\textstyle n}}  \sigma (q)
 \left[ ( \partial {\bar {\partial} } F )^n \right] \, d F
\label{poincare}
\end{equation}
where $\sigma$ satisfies eq.(\ref{defsig}) again.

  There are infinitely many symplectic 2-forms,
compatible Hamiltonian operators  and conserved
Hamiltonians for the geodesic flow for CHMA.

\section{Conclusion}

    We have considered the multi-Hamiltonian structure of various
real homogeneous Monge-Amp\`ere equations and found that quite
generally they admit {\it infinitely many} such structures.
In particular for RHMA$_2$ we have shown that there exist 
infinitely many Hamiltonian operators of both the local and 
nonlocal variety. The simplest Hamiltonian operator of the
latter type leads to the Kac-Moody algebra of vector fields and
functions on the unit circle. For Ur-RHMA$_2$ we have the scalar
version of the RHMA$_2$ Hamiltonian operators.
Finally, we have shown that local Hamiltonian operators are generic to
RHMA$_n$. Thus the real homogeneous Monge-Amp\`ere equation
is a system with infinitely many Hamiltonian, or
symplectic structures in the theory of integrable systems
{\it in arbitrary dimension}.

\section{Acknowledgement}

I am indebted to C. A. P. Galv\~{a}o for his remark on the Poisson bracket
of the constraints in Dirac's theory.
Interesting conversations with J. Swank were most useful in clarifying
the Kac-Moody algebra. I thank \"{O}. Sar{\i}o\u{g}lu for
independently checking the Jacobi identities.

\end{document}